\documentclass{svproc}

\usepackage{url}

\usepackage{natbib}
\bibpunct{(}{)}{;}{a}{}{,}%
\usepackage{graphicx}
\usepackage{amsmath,amssymb}
\usepackage{multirow}
\usepackage{longtable}

\usepackage[b5paper]{geometry}
\begin{document}
	\mainmatter              
	\title{Observations of lenticular galaxies at the 6m telescope of the Special Astrophysical Observatory}
	\titlerunning{S0s at the BTA}  
	%
	\author{Olga K. Sil'chenko\inst{1}}
	\authorrunning{Olga Sil'chenko} 
	%
	\tocauthor{Olga K. Sil'chenko}
	\institute{Sternberg Astronomical Institute of the Lomonosov Moscow State University, Moscow, Russia,\\
		\email{olga@sai.msu.su}
		}
	
	\maketitle

\begin{abstract}
        This is a historical review covering the last 30 years of the observational study of lenticular
        galaxies at the 6m telescope BTA of the Special Astrophysical Observatory of the Russian Academy of
        Sciences. The development of spectroscopic techniques at the BTA has allowed to get comprehensive
        information about this class of stellar systems, starting from the study of their nuclei in the late 80th
        towards quite exclusive results obtained in the last years on the outermost parts of their large-scale stellar disks.
\end{abstract}

\section{Introduction}

Lenticular galaxies are early-type disk galaxies; in the Local Universe according to \cite{apm} they constitute
about 15\%\ of all non-dwarf galaxies  being the second common morphological type of galaxies after spirals.
They inhabit all types of environment, from the densest ones to the quite rarified, though in clusters they are the
dominant galaxy population. High surface brightness of their bulges and disks allows to study
stellar kinematics as well as the properties of the stellar populations; however, ionized gas is also
the frequent contributor in S0s and can be studied through the analysis of optical emission lines.

Observations of lenticular galaxies at the 6m telescope had started with the emergence of digital panoramic
detectors coupled with effective spectrographs. Distinctive epochs of the S0 spectral studies are related to
the 1024-channel TV scanner of the 6m telescope, to the Multi-Pupil Fiber/Field Spectrograph (MPFS), and to the
multi-mode reducers Scorpio and Scorpio-2. The short description of the MPFS could be met in the talk by Afanasiev
with co-authors in 2001 (\cite{mpfs}), and the comprehensive descriptions of the Scorpio and Scorpio--2 were published
by V.\,L. Afanasiev and A.\,V. Moiseev (\cite{scorpio1,scorpio2}). The precious contribution to the panoramic spectral
data reduction methods was made in early 90th by \cite{vlasyuk_software}.

\section{Young stellar nuclei in lenticular galaxies}

In 1988--1989 I have succeeded to obtain spectra for the stellar nuclei of 100 nearby galaxies of different
morphological types. The observational program was fulfiled with the 1024-channel television scanner. Though
the detector used was a panoramic TV photon counter, this was in fact aperture spectroscopy because only two
4--arcsec 'strobes' were cut at the detector to read out -- one for the object and one for the sky background.
The detector was sensitive in the blue and in the extreme blue, so a set of absorption lines for which I had
measured equivalent widths included even CaIIH and K. The atlas of the spectra and the lists of the
absorption lines measured were published in 1989--1990 (\cite{scaner_1,scaner_2}).

The first look at these spectra revealed that the stellar populations of the galactic nuclei were of very
different ages, and the statistics of their ages was related to the morphological types of the host galaxies. The nuclei
in elliptical galaxies were old in the most cases. The nuclei of Sc galaxies were always young, and in the half
of them strong H$\alpha$ emission betrayed current star formation. As for the lenticulars, they
demonstrated the age distribution quite similar to that of Sa--Sb spirals and very dissimilar to that of
ellipticals. In the half of S0s strong Balmer absorpion lines were detected: the criterium of $EW(H\delta)>3$~\AA\
gave the evidence in the favour of the mean stellar age less than 3~Gyr. Now such intermediate-age galaxies
are called '$K+A$'. But the fact that the half of nearby S0s are '$K+A$' was firstly presented by \cite{scaner_3}.

Later this work was continued with the integral-field unit (IFU) MPFS: the panoramic spectroscopy allowed
to compare the ages of the nucleus and bulge stellar populations and to demonstrate distinct evolution of the stellar nuclei
within the galaxies. The particular examples of the young stellar nuclei within lenticular galaxies studied with the MPFS
were published in the series of our papers, \cite{young_s0_1,young_s0_2,young_s0_3}.

\section{Chemically decoupled stellar nuclei}

The fact that galactic stellar nuclei have quite distinct evolution with respect to the other galactic substructures
has one more bright demonstration: their metallicity has occured to be much higher than the metallicity of the underlying
regions of the bulges. We have discovered this effect during our first observations with the MPFS: by selecting in an arbitrary
way 12 luminous nearby galaxies for the test observations with the IFU, we, \cite{decnuc}, have found sharp breaks
of the MgI$\lambda$5175 and FeI$\lambda$5270,5335 radial profiles just beyond the unresolved galactic nuclei
in 7 galaxies of 12. We have called it 'chemically decoupled nuclei'; their origin has to be related obviously to
secondary star formation bursts in the galactic nuclei, sometimes quite recent. So the presence of chemically decoupled
nuclei has to be expected first of all in the 'young stellar nuclei' of lenticular galaxies.

By re-observing some of our first findings later with advanced versions of the MPFS, we confirmed that the
unresolved, point-like nuclei in the most lenticular galaxies were outstanding at the metallicity maps against their bulges
having very high stellar metallicity -- up to 2 or even 5 solar values (\cite{decnuc_mpfs1,young_s0_1,decnuc_mpfs2}).
After fifteen years of the observations of lenticulars galaxies with the MPFS/BTA I have acquired metallicity and age
maps for the central regions of 60 nearby lenticulars (\cite{decnuc_mpfs_fin}).  For this sample the nucleus age distribution
looks rather flat between $T=1$~Gyr and $T=6$~Gyr; but the age distribution of the subsample of chemically decoupled nuclei
has a prominent peak around $T=3$~Gyr delineating the distinct epoch of secondary star formation bursts in the centers of
S0 galaxies at $z\approx 0.4$ related perhaps to their assembly into massive groups and clusters.

\section{Circumnuclear polar gaseous disks}

Another interesting feature of the central regions of lenticular galaxies which has been inspected in detail by the means
of integral-field spectroscopy with the MPFS/BTA is the kinematics of the ionized gas within some hundreds parsec from their nuclei.
In the early 90th when current gas content of galactic disks was commonly considered as a remnant of initial gaseous protogalaxy,
just the gas kinematics in the centers of S0s had implied an idea of external origin of galactic cold gas supply (\cite{bertola92}).
Since by exploring the MPFS we have acquired two-dimensional distributions of many spectral parameters for the sample of nearby S0s,
including stellar and gas velocity fields, we have quickly discovered a high incidence of circumnuclear {\it polar} gas rotation.
While the coincidence of the photometric and stellar-kinematics major axes proved the stellar component circular rotation in the main
symmetry planes of the galaxies, the gas-kinematics major axis was sometimes oriented orthogonally to that of the stellar component.
We stated the existence of circumnuclear polar gaseous disks in some known lenticulars -- in NGC~7280 (\cite{young_s0_2}),
in a sample of 8 S0s with the polar dust lanes (\cite{polar_s0})...

Now the paradigm has changed, and the common point of view is that {\it all} cold gas in galactic disks, including galactic
disks of spiral galaxies, is currently accreted from outside. However, just lenticular galaxies often demonstrate decoupled rotation
of their stars and ionized gas -- according to my estimates, \cite{polar_s0_sau}, 10\%\ of nearby ($D<42$~Mpc) S0s have circumnuclear
polar gaseous disks. So the cold gas in lenticular galaxies is probably accreted along quite varied directions.

\section{Large-scale gaseous disks of lenticular galaxies}

The gas content of lenticular galaxies is not restricted only to their central regions. Many S0s -- and in
rarified environments it is the majority of them -- possess extended gaseous disks, sometimes expanding beyond the borders
of their stellar disks. We can observe them through spectroscopy of optical emission lines producing by warm ionized
gas. What we need for this task, it is spectroscopic facilities with rather large field of view. Among the
equipment of the 6m telescope, the required possibilities are provided by the long-slit mode and by the Fabry-Perot mode of
the reducers Scorpio and Scorpio-2. The field of view of the reducers is $6.1\times 6.1$~arcmin; the former mode
provides one-dimensional spectral cut-offs with a full optical spectral range covered by a single exposure while
the latter, Fabry-Perot, mode gives precise velocity measurements related to a single selected emission line over
2D field of view providing so full velocity maps for the ionized gas of the galactic disks.

Our approaches to the kinematical study of extended gaseous disks in lenticular galaxies gave many findings of
decoupled gas rotation as expected. By exploring long-slit observations, \cite{counter2} reported two S0 galaxies
with counterrotating extended gaseous disks; one of them, in NGC~5631, was inclined to the main symmetry plane
of the stellar disk and contained also a minor old stellar component. In the lenticular galaxy IC~719 studied by \cite{counter3},
the gas counterrotating the main stellar disk in its plane produced currently young stars in the ring at the fixed
radius of 1.5~kpc promising a counterrotating coplanar {\it stellar} disk in the future. Taking in mind that rarified
environments are favorable for gas kinematics decoupling, we have studied a sample of quite isolated lenticular galaxies.
\cite{counter4} reported that the majority of them possess extended ionized-gas disks, and among those, a half demonstrate
counterrotation with respect to stars when probing by a long slit aligned with the photometric major axis (the line of nodes).

However the spectral observations in long-slit mode do not provide exact information about the orientation of the
gas rotation plane: by using it we can see only the projection of the spin. When having made
for the S0 galaxy NGC~7743 several cut--offs with the long slit, we, \cite{n7743}, ensured that the line of nodes of
its gaseous disk did not coincide with the stellar disk one. To determine the fair spatial orientation of the rotation planes,
we needed 2D line-of-sight velocity distributions, and the Fabry-Perot mode was perfect for this task. Recently we have combined
the long-slit and the Fabry-Perot observations for a sample of 18 lenticular galaxies, to estimate simultaneously the
gas excitation mechanism, metallicity, and the orientation of the gas rotation plane. The results presented by \cite{s0_fabry}
are quite interesting: a quarter of our sample demonstrate the gas disk planes strongly inclined to the stellar disk over
their full extension, and just these inclined gaseous disks are excited by shocks, not by
current star formation. Even if the gas counterrotates the stars, the star formation proceeds when it is confined to the stellar
disk plane (e. g. in NGC~2551), and does not proceed when the counterrotating gas rotates in inclined plane (NGC~4143, \cite{counter6}).
Perhaps, the very origin of the S0 morphological type is related to typical inclined directions of the outer cold gas inflow
which are misaligned with the galaxy disk plane; subsequent gas heating by shocks prevents further star formation ignition in the disk.

\section{Large-scale stellar disks of lenticular galaxies}

By using spectral data obtained with the Scorpio and Scorpio--2 in the long-slit mode we have also studied large-scale
{\it stellar} disks of nearby lenticulars. As for their kinematics, we found a counterrotating extended stellar component
in NGC~448 (\cite{counter5}) and signatures of non-circular stellar rotation in the outer part of the oval disk in NGC~502
(\cite{outerdisks_kin}). But the most interesting and exclusive results were obtained for the stellar population
properties of the lenticular galaxies in a range of environments.

There exists a very popular scenario for the origin of lenticular galaxies that they are former spirals that have been devoid of
gas by tidal or ram pressure effects which are especially effective in dense environments such as clusters and groups.
This scenario was strengthened by the results of studying morphological type fraction in galaxy clusters at redshifts up to 0.8
with the Hubble Space Telescope (\cite{fasano00}): at $z>0.4$ the fraction of S0s fell at the expense of spirals so one
could conclude that just at $z=0.4$, or 4~Gyr ago, spirals coming into the clusters transformed into lenticulars. But if so, then
stellar disks of present-day lenticulars must be rather young because only 4~Gyr ago they formed stars being spirals.
We have decided to check this prediction with the long-slit spectrographs of the 6m telescope. A sample of nearby lenticulars,
firstly in groups and a few in the Virgo cluster, has been observed. \cite{outerdisks} have measured the mean ages and
metallicities of their large-scale stellar disks through the evolutionary synthesis of the Lick indices. More than 60\%\ of
the disks have revealed the ages larger than 10~Gyr! It means that the disks of nearby lenticulars stopped their star formation
at $z\approx 2$. They could not be spirals at $z=0.5$. Later this result has been confirmed by \cite{evelyn}: by studying
the sample of 13 S0s, the Virgo cluster members, they have found that {\it all} lenticulars studied have the large-scale
stellar disks older than 10~Gyr. Meanwhile the bulges look much younger than the disks. Perhaps there are just the starforming
bulges that could be taken by \cite{fasano00} as spirals in clusters at $z>0.4$.

Later \cite{stelpop_iso} have added to this sample the observations of a dozen of quite isolated S0s. Opposite to lenticulars
in dense environments, the isolated S0s have revealed the ages of the disks flatly distributed
from 1~Gyr to 15~Gyr, and their bulges are coeval with the disks. Being impressed by these results, we have proposed an alternative
scenario of the lenticular galaxy origin. All galaxies formed their thick stellar disks at $z=2$, or 10~Gyr ago, in a
brief effective star formation event, and became then lenticulars. Our Milky Way was a S0 only 10~Gyr ago! Later many of them
have accreted cold gas from outside and have re-commenced star formation under favorable orientation of the accretion flows.
But S0s which had fallen into clusters had no possibility of cold gas inflow due to hot intergalactic gas ram pressure and
strangulation; it is because there are now so many S0s in clusters.

\section{Outer starforming rings in S0s}

Last few years I become interested in the outermost parts of the S0 disks -- their outer starforming rings. Just in lenticular
galaxies the outer rings are very frequent -- more than 50\%\ S0s have outer stellar rings (\cite{arrakis}), -- while the
bar fraction drops in S0s related to spirals (\cite{lauri}). It means that outer {\it gaseous} rings in S0s may be mostly formed
by outer gas accretion, and by studying gas characteristics in the rings we can restrict its origin and say something
about cold-gas accretion sources.

After the GALEX discovery of UV-bright rings in early-type galaxies (\cite{marino}) we started their spectral study at the 6m telescope
(\cite{rings1}). Many interesting particular cases have been analysed (\cite{rings2,rings3,rings4}) including both resonant rings
and rings produced by obvious accretion from outside. For the rings confined to the galactic planes we fixed normal, for their gas
content, star formation rates and nearly solar metallicity (\cite{s0_fabry}). The latter fact seems to restrict possible sources
of outer gas accretion to rather large gas-rich satellite tidal disruption.

\paragraph{Acknowledgements.}
This work is based on the spectral data obtained at the Russian 6m telescope of the Special Astrophysical Observatory
carried out under the financial support of the Ministry of Education and Science of the Russian Federation
(agreement No05.619.21.0016, project ID RFMEFI61919X0016). The study of the outer rings in S0s is supported by the
grant of the Russian Foundation for Basic Researches, RFFI 18-02-00094a.

%
%
%
%
%
%
%
%
%
%

\bibliographystyle{aa}
\bibliography{silchenko}
\end{document}